\documentclass[journal=jacsat,manuscript=article]{achemso}

\usepackage[version=3]{mhchem} 
\usepackage{xcolor}
\newcommand{\red}{\textcolor{black}}
\newcommand{\RV}[1]{{\red{{\bf} #1}}}

\author{Valeriy I. Kondratyev}
\affiliation{School of Physics and Engineering, ITMO University, St. Petersburg, 197101, Russia}
\email{valeriy.kondratiev@metalab.ifmo.ru}
\author{Dmitry V. Permyakov}
\author{Tatyana V. Ivanova}
\author{Ivan V. Iorsh}
\affiliation{School of Physics and Engineering, ITMO University, St. Petersburg, 197101, Russia}
\author{Dmitry N. Krizhanovskii}
\affiliation
{Department of Physics and Astronomy, University of Sheffield, Sheffield S3 7RH, UK}
\author{Maurice S. Skolnick}
\affiliation
{Department of Physics and Astronomy, University of Sheffield, Sheffield S3 7RH, UK}
\author{Vasily~Kravtsov}
\affiliation{School of Physics and Engineering, ITMO University, St. Petersburg, 197101, Russia}
\author{Anton K. Samusev}
\affiliation{Experimentelle Physik 2, Technische Universit\"at Dortmund, 44227 Dortmund, Germany}
\alsoaffiliation{School of Physics and Engineering, ITMO University, St. Petersburg, 197101, Russia}
\email{anton.samusev@gmail.com}

\title{Probing and control of guided exciton-polaritons in~a~2D~semiconductor-integrated slab waveguide}

\abbreviations{TMD, SIL, BFP}


\makeatletter
\newcommand*{\forcekeywords}{
  \acs@keywords@print
  \let\acs@keywords@print\relax
}
\makeatother

\begin{document}

\begin{tocentry}

\begin{center}
    \includegraphics[width=\textwidth]{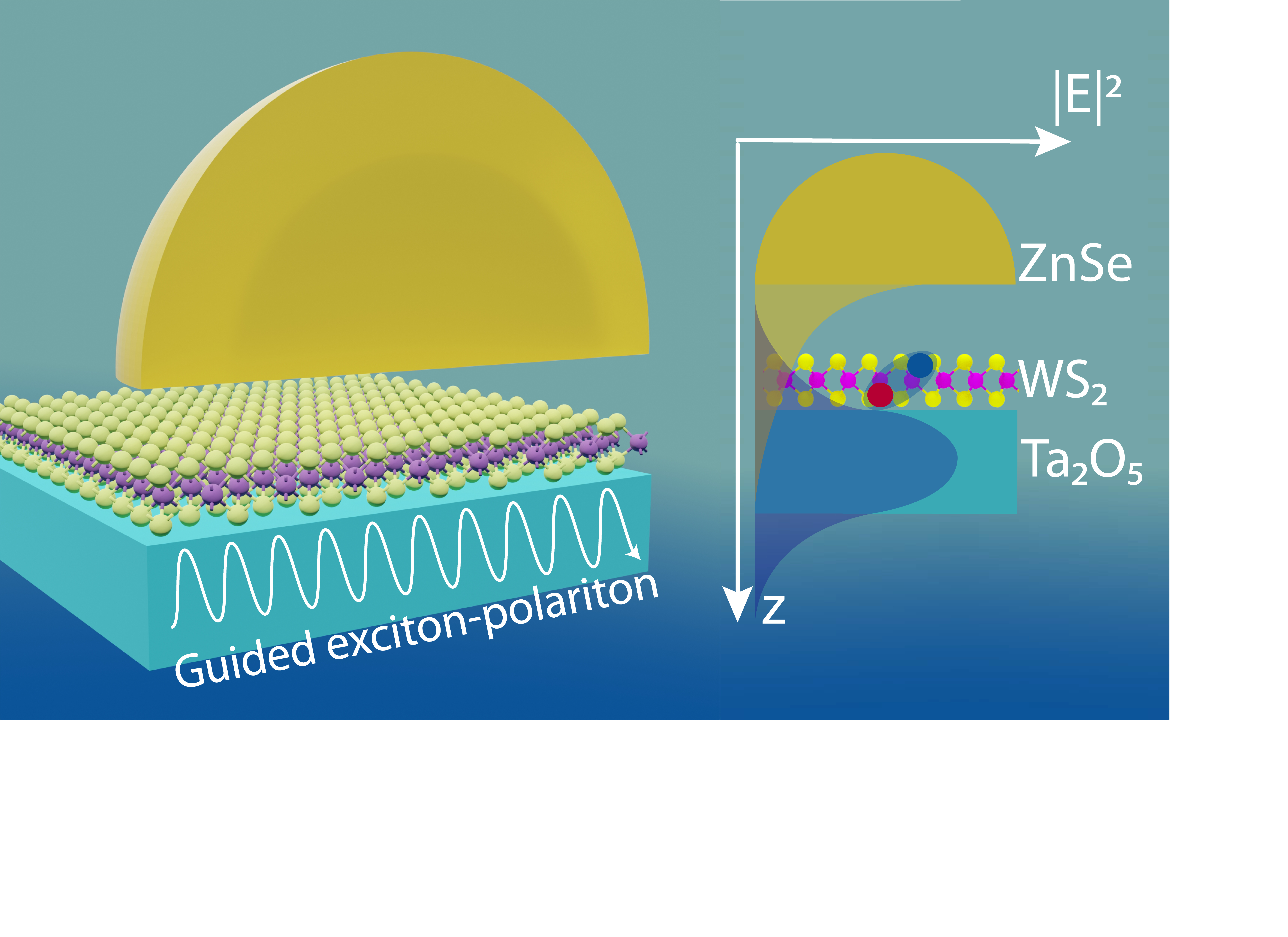}
\end{center}
    
\end{tocentry}

\begin{abstract}
Guided 2D exciton-polaritons, resulting from the strong coupling of excitons in semiconductors with non-radiating waveguide modes, provide an attractive approach towards developing novel on-chip optical devices. These quasiparticles are characterized by long propagation distances and efficient nonlinear interaction but cannot be directly accessed from the free space. Here we demonstrate a powerful approach for probing and manipulating guided polaritons in a Ta$_2$O$_5$ slab integrated with a WS$_2$ monolayer using evanescent coupling through a high-index solid immersion lens. Tuning the nanoscale lens-sample gap allows for extracting all the intrinsic parameters of the system. We also demonstrate the transition from weak to strong coupling accompanied by the onset of the motional narrowing effect: with the increase of exciton-photon coupling strength, the inhomogeneous contribution to polariton linewidth, inherited from the exciton resonance, becomes fully lifted. Our results enable the development of integrated optics employing room-temperature exciton-polaritons in 2D semiconductor-based structures.

\end{abstract}

\forcekeywords

Polaritonics has attracted substantial attention as a promising base for devices utilizing non-linear effects. Polaritons arise from strong coupling between light and resonant transition in matter and manifest themselves in the energy spectrum as Rabi splitting~\cite{khitrova2006}. One of the promising materials for polaritonics is transition metal dichalcogenides (TMDs) \cite{Schneider2018}. In their monolayer limit, TMDs are direct bandgap semiconductors \cite{mak2016photonics}, and their optical response is dominated by the excitonic resonance\cite{wang2018colloquium}. TMD excitons possess high oscillator strength with large binding energy, and they are stable in ambient conditions \cite{chernikov2014exciton}, which makes these materials ideal candidates for room-temperature polaritonic devices \cite{gu2019room}.

The strong light--matter coupling regime with exciton-polariton formation can be achieved by coupling excitonic resonances in TMD monolayers to confined optical modes supported by distributed Bragg reflector (DBR) mirrors\cite{dufferwiel2015exciton,dufferwiel2018valley,wurdack2021motional,zhao2021ultralow}, plasmonic structures \cite{sun2021strong,jiang2019tunable,liu2016strong}, or subwavelength gratings\cite{kravtsov2020nonlinear,zhang2018photonic}.
However, the polariton propagation length in such structures is limited by their radiative decay rates and small group velocity.
In contrast, guided polaritons~\cite{beggs2005waveguide, walker2013exciton}, where excitons couple to non-radiating waveguide modes with large group velocity, are characterized by long-range propagation up to hundreds of $\mu m$~\cite{liran2018fully}, which is very useful for the integration in photonic circuits.
This, combined with sizeable polariton--polariton interaction, leads to a number of technologically important non-linear optical effects such as soliton formation~\cite{walker2019spatiotemporal,walker2015ultra}, lasing~\cite{suarez2020electrically}, and self-phase modulation~\cite{walker2019spatiotemporal}.
In addition, unpatterned waveguide structures supporting guided polaritons require far fewer and significantly cheaper fabrication processes.

Exciton-polaritons in planar unpatterned structures have been studied in different material platforms~\cite{pirotta2014strong,liran2018fully} including bulk TMD crystals~\cite{munkhbat2018self,gogna2020self}.
Due to the non-radiating nature of guided polaritons, probing their dispersion in experiment is challenging.
One approach is provided by scattering-type scanning near-field optical microscopy~\cite{hu2019imaging, iyer2022nano, luan2022imaging}.
\red{While it offers nanoscale lateral spatial resolution and provides access to the information on phase and propagation lenghts of the waveguide modes}, it requires complex and time-consuming measurements based on raster-scanning the sample with a near-field probe \red{and does not allow to controllably tune the parameters of the system under study}.

Another approach relies on the use of high-NA oil-immersion objectives~\cite{shin2022direct, munkhbat2018self,canales2023perfect}.
It allows imaging of the polariton dispersion in a single optical measurement but does not provide means for controlling the parameters of the strongly coupled system and is limited to room conditions due to the use of immersion oil.

In this work, we probe and control guided exciton-polaritons in monolayer WS$_2$ integrated with a 90-nm-thick Ta$_2$O$_5$ waveguide using the approach based on attenuated total internal reflection~\cite{otto1968excitation}.
To excite and detect the polaritons, which are intrinsically uncoupled from free-space waves and propagate below the light line, we use Fourier-plane microscopy with a solid immersion lens (SIL)~\cite{permyakov2021probing}.
The use of SIL with an accurately controlled distance to the sample surface allows us not only to probe the polariton dispersion but also to tune the radiative losses of the waveguide mode, thus controlling the regime of strong coupling.
\RV{Notably, this technique does not require the use of edges/defects to create standing polariton waves and therefore provides a powerful microscopy tool for a wide range of samples, which is compatible with various advanced optical spectroscopic methods, e.g., nonlinear optical spectroscopy based on harmonics generation, parametric down conversion, or four-wave mixing processes.}
Our results establish a novel approach to study and control guided exciton-polaritons in 2D semiconductors and provide a basis for future investigations of radiative/non-radiative losses, lifetimes, and nonlinearities in systems supporting guided polaritons.


\begin{figure}
\includegraphics[width=\textwidth]{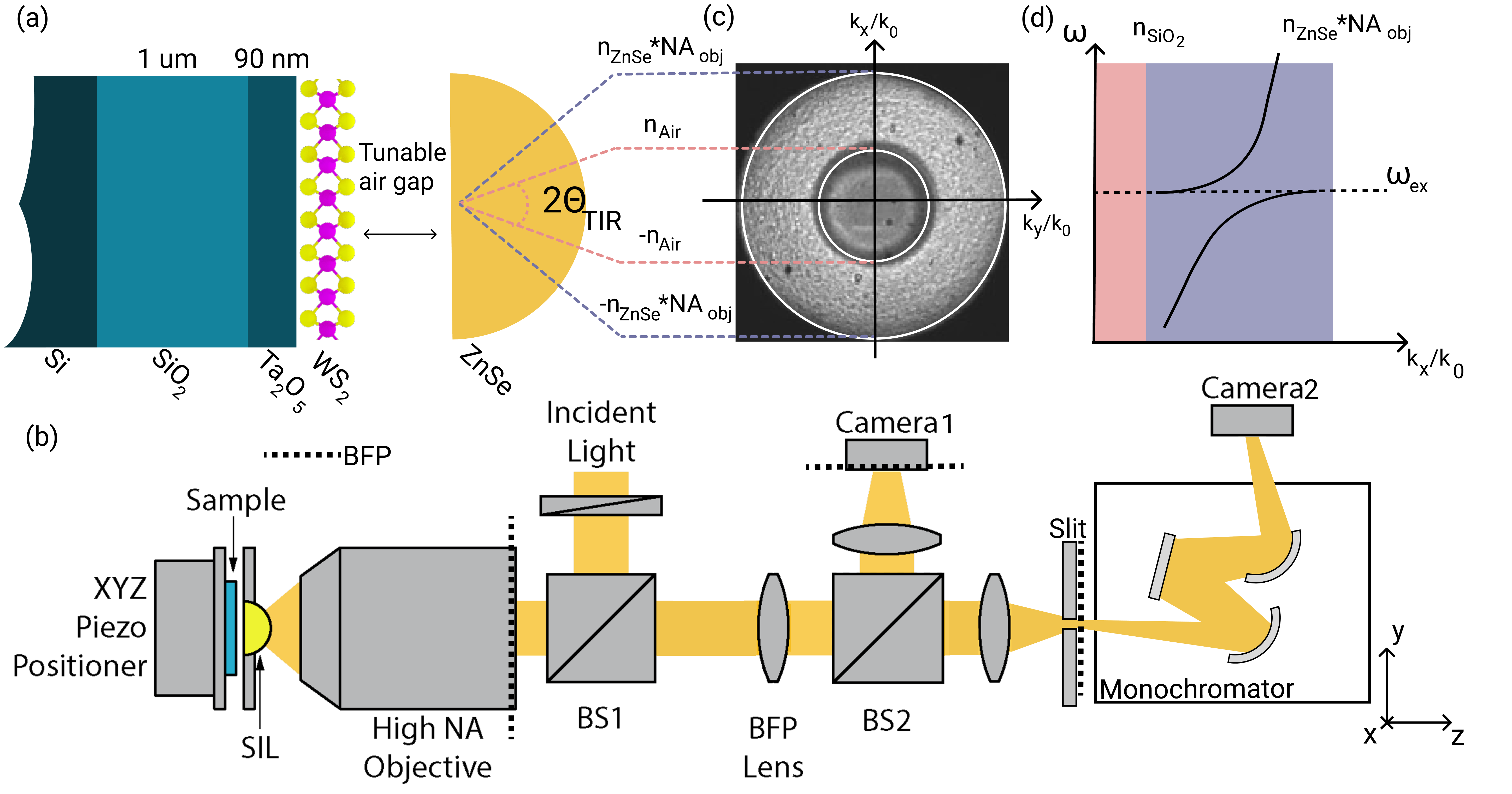}
\caption{(a) Schematic of planar Ta$_2$O$_5$ optical waveguide on SiO$_2$/Si substrate coupled to a WS$_2$ monolayer. ZnSe hemisphere is used as a solid immersion lens (SIL) attached to the sample with a 100-nm-scale separation gap which can be tuned in the experiment. The size of the gap allows precise control of the radiative losses of the optical modes supported by the studied system. (b) The layout of the back focal plane microscopy and spectroscopy setup combined with a SIL.
(c) Angle-resolved white light reflectance map in BFP obtained by Camera 1. The inner region corresponds to in-plane wavevectors smaller than $k_0 \cdot n_\mathrm{Air}$, and the outer region represents states with $k_0 \cdot n_\mathrm{Air} < k < k_0 \cdot n_\mathrm{ZnSe} \cdot \mathrm{NA}_\mathrm{obj}$, where $n_\mathrm{ZnSe} \cdot \mathrm{NA}_\mathrm{obj}$ is the effective numerical aperture in the setup. (d) Sketch of a typical guided polariton dispersion, which can be measured in our experimental scheme.}
\centering
\label{fig:SampleAndSetup}
\end{figure}

The system under study is shown in Figure~\ref{fig:SampleAndSetup}(a). WS$_2$ monolayer was mechanically exfoliated from bulk crystal and dry transferred on 90-nm-thick Ta$_2$O$_5$ optical single-mode waveguide on Si/SiO$_2$ substrate, with the thickness of SiO$_2$ spacer of 1~$\mu m$. \red{We note that the waveguide thickness is chosen such that only the fundamental guided TE mode is supported in the spectral range of interest (see Supplementary Materials Figure~S12).} We used WS$_2$ since it exhibits the most favorable ratio of the oscillator strength to exciton linewidth for achieving strong exciton--photon coupling at room temperature.
The experimental scheme is shown in Figure~\ref{fig:SampleAndSetup}(b). For excitation and detection of guided polaritons with large wavevectors, we combine evanescent wave coupling through SIL with a custom-built back focal plane (BFP) spectroscopy setup.

In the BFP spectroscopy part, the light coming from the beam splitter (BS1) is focused onto the sample by a microscope objective with a large numerical aperture (Mitutoyo, M Plan Apo HR, 100×, NA = 0.9). The reflected light passes through the 4f scheme, and the BFP of the objective is imaged onto a spectrometer slit coupled to a liquid-nitrogen-cooled imaging CCD Camera~2 (Princeton Instruments SP2500 + PyLoN), allowing us to observe reflection as a function of both wavevector and frequency. The slit cuts the BFP image in the $k_x/k_0$ direction at $k_y = 0$. Beamsplitter BS2 directs a part of the reflected light to Camera 1 (CMOSIS CMV300 USB3), which allows us to simultaneously observe the BFP image from the sample as shown in Figure~\ref{fig:SampleAndSetup}(c) and capture dispersions with Camera 2 as schematically shown in Figure~\ref{fig:SampleAndSetup}(d). We use an analyzer in the detection path to probe only TE waveguide modes for all reported measurements.

The in-plane wavevectors that can be accessed in an optical microscopy setup with a dry objective lens are limited by the numerical aperture, which does not exceed unity: $k_{||}/k_0 \leq \mathrm{NA} < 1$. To  overcome this limitation, we introduce a hemispherical SIL made of ZnSe in the optical path and use attenuated total internal reflection (TIR) to excite guided modes (characterized by larger wavevectors) in our sample evanescently. Figure~\ref{fig:SampleAndSetup}(c) shows the BFP reflectance map obtained from a bare ZnSe lens. The light reflected from the center of the flat side of the lens can be separated into two regions: (i) with angles smaller than TIR angle shown in Figure~\ref{fig:SampleAndSetup}(c) as the inner circle and (ii) with angles larger than TIR angle (outer circle). The inner region corresponds to $k_{x}/{k_0} < n_\mathrm{Air}$, where $n_\mathrm{Air}$ is the air refractive index, as bulk waves exist in the air and manifest themselves as a dip in the center of Figure~\ref{fig:SampleAndSetup}(c). In the outer circle  $k_{x}/{k_0} > n_\mathrm{Air}$, the waves propagating below the light line can exist in the air as evanescent waves. The resulting effective numerical aperture in our experiment is
\begin{equation}
    \left(\frac{k_x}{k_0}\right)_{max}=\mathrm{NA}_\mathrm{eff}=\mathrm{NA}_\mathrm{obj} \cdot n_\mathrm{SIL} \approx 2.25,
\end{equation}
where $k_0=\omega / c$ is the free space wavevector of light, $c$ and $\omega$ are speed and frequency of light,  and n$_\mathrm{SIL}$ stands for the refractive index of SIL. In our case $n_\mathrm{SIL} = n_\mathrm{ZnSe} \approx 2.5$, $\mathrm{NA}_\mathrm{obj} = 0.9$. Note that one can use a different material for SIL such as GaP, with refractive index $\approx 3.5$, to achieve an even higher effective NA~$\approx$~3. The substrate of our sample has a 1~$\mu m$ SiO$_2$ spacer layer, so the states above the SiO$_2$ light line leak to the substrate. We outline it with red color in Figure~\ref{fig:SampleAndSetup}(d). The states with $k_x/k_0 > n_\mathrm{SiO_2}$, where $n_\mathrm{SiO_2} \approx 1.45$, are completely decoupled from the free space radiation in both substrate and air.



\begin{figure}
\includegraphics[width=\textwidth]{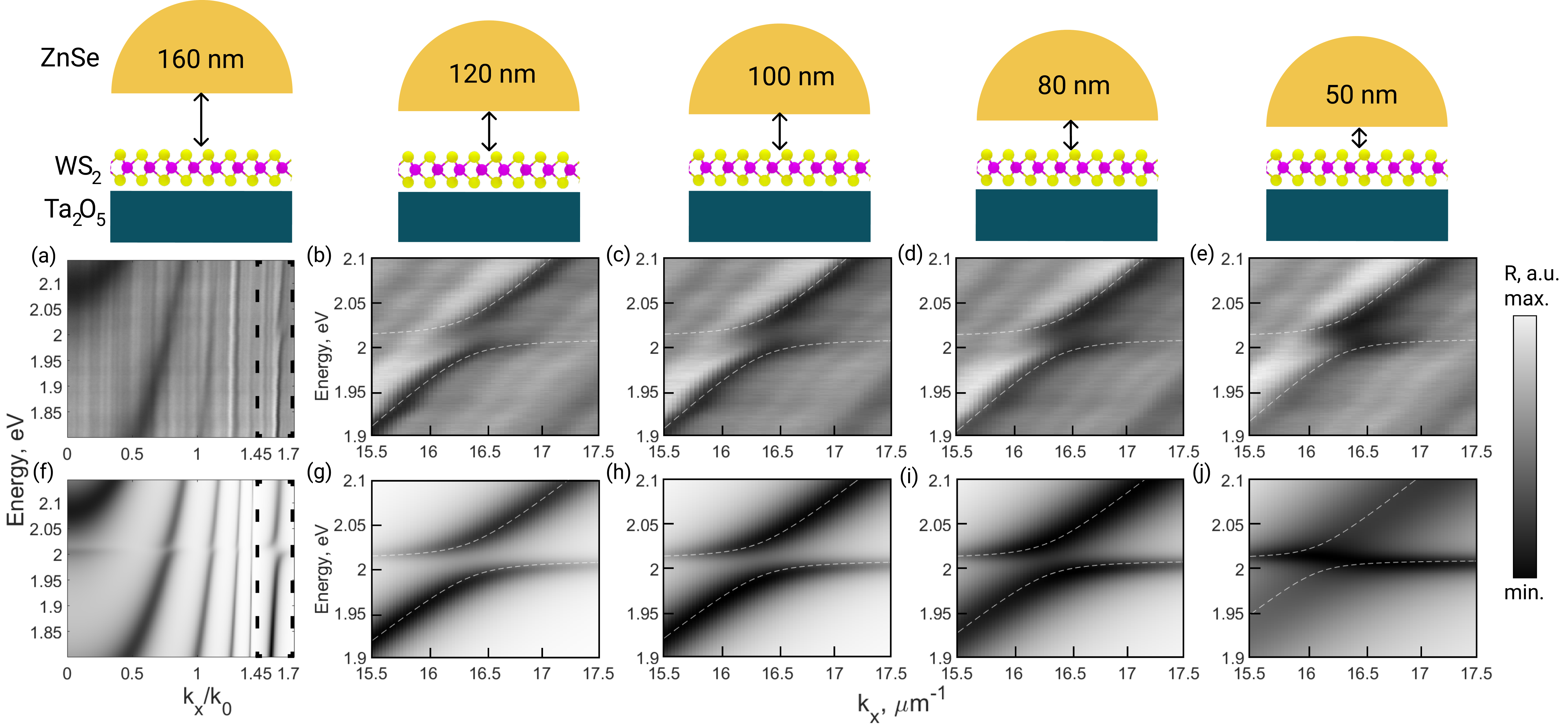}
\caption{(a)-(e) Measured and (f)-(j) numerically calculated angle-resolved reflectance maps for different air gaps. The region framed with black dashed lines in (a) and (f) is magnified in (b)-(e) and (g)-(j). Schemes in the top row show the SIL-sample distance (air gap) for each pair of experimental and simulated maps. The white dashed line shows the dispersion fitted using the model of two coupled oscillators.}
\centering
\label{fig:GapsAndMaps}
\end{figure}

In the experiment, the sample is attached to a 3-axis piezo positioner. Moving along the $x$ and $y$ axes allows positioning the focal spot on a TMD monolayer, while moving in the z direction controls the SIL--sample gap. Figure~\ref{fig:GapsAndMaps}(a-e) shows measured angle-resolved reflectance maps for different SIL--sample gaps as sketched in the top row. 
The interference fringes observed at $k_x/k_0 < n_\mathrm{SiO_2}$ in Figure~\ref{fig:GapsAndMaps}(a),(f) correspond to the Fabry-Perot modes \red{arising due to the presence of} the SiO$_2$ layer.
\red{Since the spectral positions of these fringes show a near-linear and sensitive dependence on the SIL--sample distance, we employ the comparison between measured and simulated normal-incidence reflectance spectra for calibrating the air gap values in the experimental data to best fit the dependencies in Figure~\ref{fig:ReflectanceLossesAndRabi} (a,b). The details on air gap calibration are given in Supplementary Materials Section 3.} 

In the measured \red{angle-resolved reflectance maps shown in Figure~\ref{fig:GapsAndMaps}(a-e)}, one can see the waveguide mode in the Ta$_2$O$_5$ layer at $k_x/k> n_\mathrm{SiO2}$ as highlighted with a dashed rectangle (see also Supplementary Materials Figure S1). The panels in Figure~\ref{fig:GapsAndMaps}(b)-(e) show zoomed-in dispersion regions at the splitting between the guided mode and WS$_2$ monolayer exciton resonance at the energy of $\approx$ 2.01 eV. 

To support our experimental data, we perform numerical simulations using the T-matrix method~\cite{waterman1965matrix}. For Ta$_2$O$_5$ waveguide layer, we use refractive index data from Ref.~\citenum{rodriguez2016self} with increased losses fitted to account for additional scattering induced by the monolayer and SIL inhomogeneities. For excitons in WS$_2$ monolayer, we use  $\gamma_{exc}^{nr}$ = 8.9~meV, $\gamma_{exc}^{r} = 1.9$~meV, and E$_{exc} = 2.01$~meV, where $\gamma_{exc}^{r}$ and $\gamma_{exc}^{nr}$ are the radiative and non-radiative decay rates of exciton resonance with a total linewidth of $\gamma^{h}_{exc}=\gamma_{exc}^{nr}+\gamma_{exc}^{r}$, and E$_{exc}$ represents the energy of the exciton resonance. Note that the non-radiative part $\gamma_{exc}^{nr}$ accounts for various effects such as phonon scattering at room temperature, phonon-induced dephasing, and non-radiative recombination.
The choice of these specific values is justified further in the text.
The results of the simulations for different air gaps are shown in Figure~\ref{fig:GapsAndMaps}(f-j).

Both experimental and simulated angle-resolved reflectance maps in Figure~\ref{fig:GapsAndMaps} show that, with the decrease of the SIL--sample gap, the waveguide mode becomes broader and the splitting between the exciton resonance and waveguide mode becomes less visible.
To extract the SIL-induced modification of the reflectance and losses of the waveguide mode, we fit the dependence of reflectance on the in-plane wavevector $k_x/k_0$ at 2.05 eV for both experimental and simulated data with Fano line shapes (see Supplementary Materials Section 1 for the details on the fitting procedure). The large detuning from the exciton resonance (2.01 eV) ensures that the mode at this wavevector is close to being purely photonic. We then extract the SIL-induced reflectance modulation and mode linewidth as the amplitude and FWHM of the fitted Fano line shape, respectively.

The obtained reflectance modulation and total waveguide losses depending on SIL--sample distance are shown with back dots in Figure~\ref{fig:ReflectanceLossesAndRabi}(a) and (b), respectively. Note that with the decrease of the SIL--sample gap, the waveguide mode radiative losses $\gamma_c^{r}$ increase, while non-radiative ones $\gamma_c^{nr}$ originating from light absorption and scattering on defects remain the same. From the reflectance spectrum obtained for a specific gap, we can only extract the total linewidth (losses) of the mode $\gamma_c = \gamma_c^{nr}+\gamma_c^{r}$, which is shown in Figure~\ref{fig:ReflectanceLossesAndRabi}(b). The gap-independent nonradiative decay rate of the mode can be estimated using two different approaches. First, it can be obtained as the asymptotic value of total losses at large SIL--sample gaps when the radiative losses become negligible (horizontal dashed line in Figure~\ref{fig:ReflectanceLossesAndRabi}(b)). 
Second, it can be estimated from the characteristic minimum of reflectance modulation at a specific gap seen in Figure~\ref{fig:ReflectanceLossesAndRabi}(a), which corresponds to the critical coupling regime \cite{permyakov2021probing} where the radiative and nonradiative losses become equal to each other: $\gamma_c^{nr} = \gamma_c^{r}$ (vertical dashed line). 
\red{
The nonradiative decay rate is then taken as half of the total linewidth at the point of critical coupling extracted from Figure~\ref{fig:ReflectanceLossesAndRabi}(b).}

\begin{figure}
\includegraphics[width=0.5\textwidth]{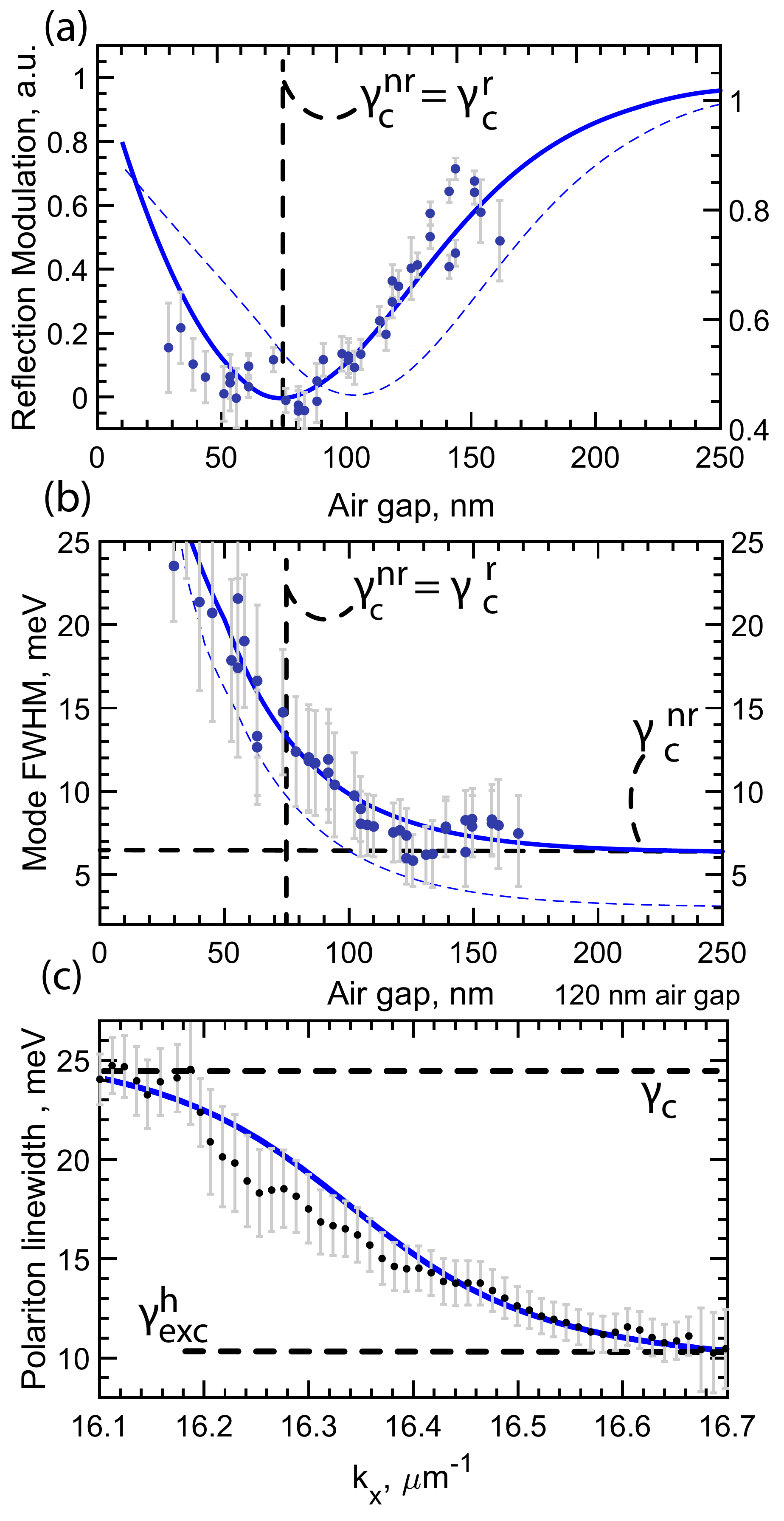}
\caption{(a) Experimental (dots) and simulated (curves) reflectance modulation as a function of SIL--sample gap size at 2.05~eV. The vertical dashed line indicates the gap corresponding to the critical coupling regime. (b) Photonic mode FWHM obtained from experiment (dots) and simulation (curves). In the critical coupling regime, the radiative and non-radiative losses equalize. Blue dashed curves in (a, b) correspond to simulations with the original losses in Ta$_2$O$_5$, and solid curves correspond to simulations with additional losses associated with light scattering on TMD imperfections. (c) Extracted polariton FWHM as a function of wavevector at $\approx 120$~nm SIL--sample gap, with experimental data (dots) and model fit (curve). The horizontal dashed lines denote the total losses of the photon mode ($\gamma_c$) and the exciton resonance ($\gamma^{h}_{exc}$).}
\centering
\label{fig:ReflectanceLossesAndRabi}
\end{figure}

Comparison between the data extracted from the experiment and numerical results suggests the presence of additional losses, which we attribute to scattering at the TMD monolayer. To account for this scattering, we increase the imaginary part of the waveguide material (Ta$_2$O$_5$) refractive index in simulations. For comparison, the simulations with the original material dispersion are shown as dashed curves in Figure~\ref{fig:ReflectanceLossesAndRabi}(a, b). 

In contrast to the case of the photonic mode, the exciton resonance properties are not affected by the SIL for the experimental range of gaps, \red{since the exciton size is at least an order of magnitude smaller than the air gaps studied in our experiment.}
Therefore, with the gaps above 10~nm, we tune only the radiative part of the waveguide mode losses, while excitonic properties remain intact. The difference in the effect of the SIL on the exciton and the waveguide mode resonances allows us to independently determine excitonic and waveguide losses from the experimental data, as discussed below. 

\begin{figure}
 \includegraphics[width=0.5\textwidth]{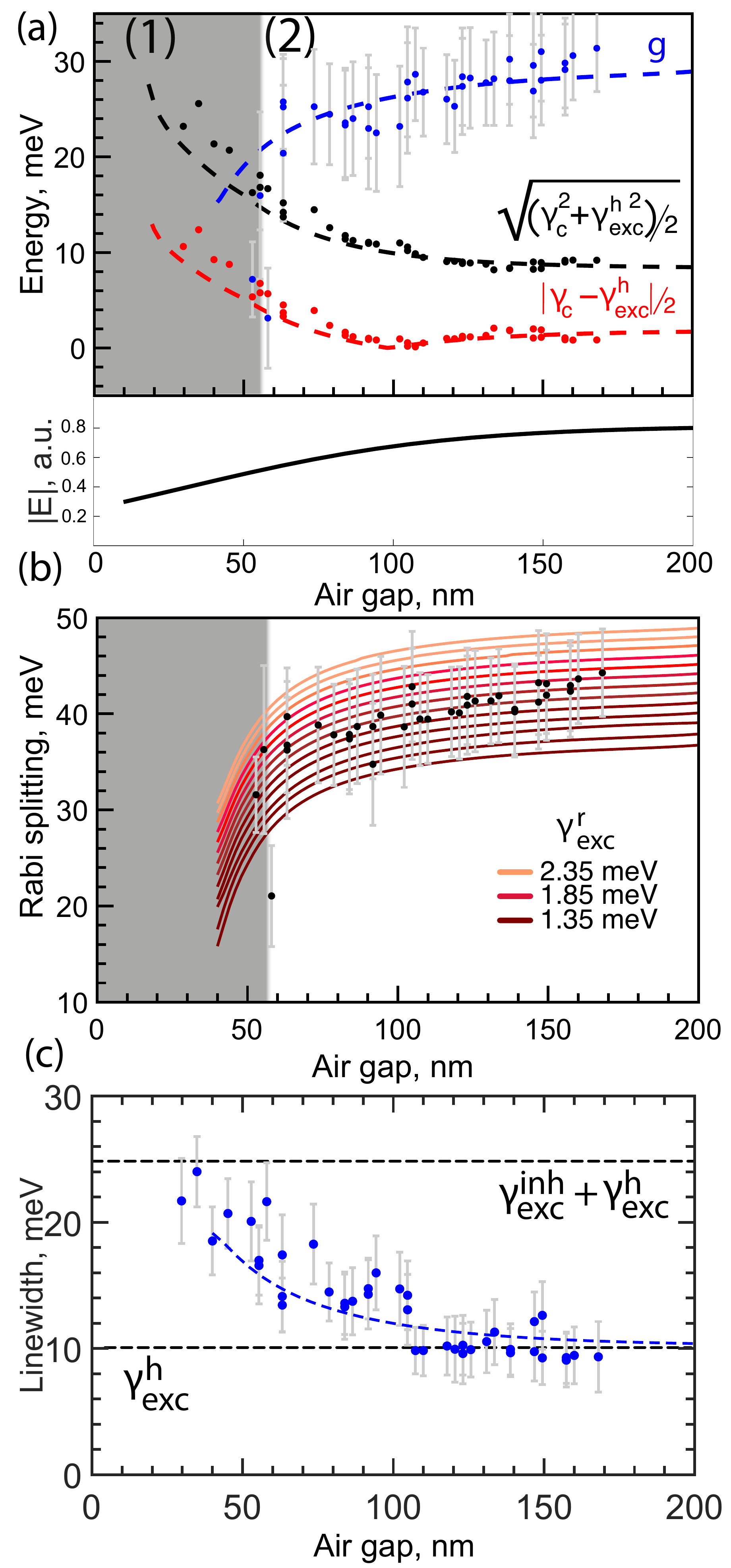}
\caption{ (a) Coupling strength (blue color) in comparison with the expressions for the two strong coupling criteria (red and black colors) obtained from the experimental (dots) and simulated (dashed line) data depending on the air gap. \red{The bottom panel shows the evolution of the electric field on top of the waveguide, where the monolayer resides}. (b) Extracted (black dots) and simulated (solid lines) Rabi splitting as a function of the air gap. Simulation results are shown for different exciton radiative decay rates. In both panels, the shaded area corresponds to the SIL--sample distance, where the system \red{undergoes the transition to the weak coupling regime.} (c) \RV{Extracted polariton linewidth at $k_x= 17\,\mu$m$^{-1}$ as a function of SIL--sample gap. 
The blue dashed line corresponds to the theoretically predicted polariton linewidth with account for the motional narrowing effect. Dashed horizontal lines represent the limiting cases of full ($\gamma^{inh}_{exc}+\gamma^{h}_{exc}$) and homogeneous ($\gamma_{exc}^{h}$) exciton linewidths.}}
\centering
\label{fig:PolaritonAndRadiativeLosses}
\end{figure}

Measured and simulated maps depending on the SIL--sample gap are shown in the Supplementary Materials video. A detailed description of the procedure we follow to fit the polariton dispersion is given in Supplementary Materials Section~2. As justified further, the waveguide mode is strongly coupled to the exciton resonance. This is manifested as Rabi splitting in the reflectance maps. One could expect the modification of the Rabi splitting with the alternation of waveguide mode losses in accordance with the relation:
\begin{equation}
   \Omega_{Rabi}=\sqrt{4g^2-(\gamma^{h}_{exc}-\gamma_c)^2}  ,
\end{equation}
where $g$ is the exciton-photon coupling strength. Dependencies of exciton-photon coupling strength and Rabi splitting on the SIL--sample gap extracted from the experimental data are shown in Figure~\ref{fig:PolaritonAndRadiativeLosses}(a) and (b). These plots can be split into two regions: (1) air gaps smaller than $\sim$ 50 nm, and (2) air gaps exceeding $\sim$ 50 nm. For larger gaps (Region~2), the field of the waveguide mode is weakly modified by SIL, and therefore the coupling strength only slightly depends on the air gap. In contrast, in Region~1, with the decrease of the SIL--sample gap, the mode field profile becomes strongly affected by the SIL, which is manifested by the drastic increase of its radiative losses.

To confirm that the studied system is in the strong coupling regime, we check its parameters against the two strong coupling criteria. According to the first criterion, the value of the Rabi splitting $\Omega_{Rabi}$ should be real, which yields $g > |(\gamma^{h}_{exc}-\gamma_{c})|/2$. In Figure~\ref{fig:PolaritonAndRadiativeLosses}(a), the left- and right-hand sides of this inequality, extracted from experimental data, are plotted with blue and red dots, respectively. 
One can see that the first criterion holds for all studied SIL--sample gaps. According to the second criterion, to spectrally resolve the upper and lower polaritonic resonances, the polariton splitting
has to be greater than the sum of the 
linewidths of the upper and lower polariton branches, which is equivalent to $g > \sqrt{(\gamma_{exc}^{h2}+\gamma_{c}^2)/2}$. The right-hand side of this inequality is shown with black color in  Figure~\ref{fig:PolaritonAndRadiativeLosses}(a).
One can see that for SIL--sample gaps below $\sim 50$~nm this condition is no longer met. This transition is also clearly seen in Figure~\ref{fig:GapsAndMaps}, where the splitting between the upper and lower polariton branches becomes indistinguishable in panels (e) and (j).

The two regions in Figure~\ref{fig:PolaritonAndRadiativeLosses}(a) and (b) therefore formally represent the strong (Region 2) and weak (Region 1, shaded area) coupling regimes. Note that in addition to the change in losses and thus in the imaginary part of the wavevector, its real part also changes for small air gaps, as shown in Supplementary Materials Figure~S4. This is a fingerprint of the alternation of the mode effective refractive index stemming from the  change in the mode field distribution in the presence of the high refractive index SIL separated from the sample by a small distance (below $\approx 50$~nm, Region 1). The two-oscillator model based on the unperturbed states we use for fitting the data becomes not applicable in this regime since the mode is not well-localized in the waveguide layer anymore.

\red{The observed evolution of exciton-photon coupling with the increase of the air gap allows us} to distinguish between the radiative and non-radiative decay rates of the exciton in the WS$_2$ monolayer. The coupling strength between the exciton and waveguide resonances is related to the exciton oscillator strength $f$ as $g \propto \sqrt{f} \cdot E$, where $E$ is the overlap integral defined by the mode profile~\cite{koshelev2018strong}. The exciton oscillator strength is, in turn, proportional to its radiative decay rate $\gamma_{exc}^{r}$.
Therefore, we can determine $\gamma_{exc}^r$ using the asymptotic value of \RV{the  Rabi splitting} for large air gaps, see Figure~\ref{fig:PolaritonAndRadiativeLosses}(b). 
Simulation results for different exciton radiative decay rates are shown in Figure~\ref{fig:PolaritonAndRadiativeLosses}(b) with solid lines.
Comparing simulated and experimental data and minimizing the root mean square error (see Supplementary Materials Figure S5), we obtain the exciton radiative losses of \RV{$\gamma_{exc}^r=1.9 $~meV}.
Since the \red{homogeneous exciton linewidth is} $\gamma_{exc}^{h} = 10.5$~meV, we extract the exciton non-radiative decay rate \RV{as $\gamma_{exc}^{nr}=8.6 $~meV.}

\red{Remarkably, the homogeneous exciton linewidth extracted from the polariton dispersion is significantly smaller than the linewidth of the uncoupled exciton~\cite{selig2016excitonic}, which in our case is obtained from PL spectra as $\gamma_{exc} = 25$~meV (see Supplementary Materials Figure S11).
This difference stems from the fact that the inhomogeneous broadening in the strong coupling regime can be lifted due to the effect of motional narrowing\cite{whittaker1996motional}. Since changing the SIL--sample air gap allows us to achieve the transition from weak to strong coupling, we can study the onset of motional narrowing in our system. To do so, we plot the polariton linewidth at $k_x= 17\,\mu$m$^{-1}$ (corresponding to a high exciton fraction) as a function of the air gap, as shown in Figure~\ref{fig:PolaritonAndRadiativeLosses}(c).
}

\red{
For large air gaps, where the exciton--photon coupling strength is high and the motional narrowing effect is most pronounced, the polariton linewidth corresponds to the homogeneous linewidth of the exciton resonance $\gamma_{exc}^{h}$. With the decrease of the gap, the coupling strength is gradually reduced, and the effect of motional narrowing eventually vanishes, such that the corresponding linewidth starts to include the inhomogeneous contribution: $\gamma_{exc} = \gamma_{exc}^{h}+\gamma_{exc}^{inh}$.
In order to describe the observed experimenal result theoretically, we use a model recently proposed in literature\cite{osipov2023transport}, as outlined in Supplementary Materials Section~2.
The theoretical result is shown in Figure~\ref{fig:PolaritonAndRadiativeLosses}(c) with a dashed blue curve and demonstrates good agreement with the experimental data.
The observed pronounced dependence of the polariton linewidth on the SIL--sample gap in the transition between strong and weak light--matter coupling is an important manifestation of the polariton motional narrowing effect.
Thus, by tuning the geometrical parameters of our experimental scheme, we demonstrate control on the motional narrowing effect, with the ability to switch on and off the inhomogeneous broadening of excitons in monolayer semiconductors.
}

\section{Conclusion}
In summary, in this work, we directly probe guided exciton-polaritons in a planar TMD-based waveguide.
To access their dispersion below the light line, we use a solid immersion lens combined with the back focal plane spectroscopy technique.
Careful measurement of angle-resolved reflectance maps as a function of SIL--sample air gap enables reversible control of the waveguide and exciton-polariton dispersions.
Our experimental results combined with numerical simulations allow us to retrieve the parameters of the system, which include: radiative and nonradiative decay rates of the exciton resonance, as well as the dependence of corresponding rates for the optical mode on the SIL--sample distance.
Additionally, we demonstrate control on exciton-polariton coupling strength via SIL-induced rearrangement of the local electromagnetic field in the polaritonic waveguide.
\red{Utilizing the transition from weak to strong coupling regime, we experimentally observe the onset of the motional narrowing effect and demonstrate control on the inhomogeneous broadening of polaritons.}
The suggested approach is suitable for probing guided polaritons in arbitrary planar structures and can be employed in various fields including active and nonlinear topological polaritonics.

\begin{acknowledgement}
Sample fabrication was supported by Priority 2030 Federal Academic Leadership Program.
Optical measurements were funded by Russian Science Foundation, project 21-72-10100.
Theoretical modelling was funded by Russian Science Foundation, project 21-12-00218.
A.K.S. acknowledges TU Dortmund core funds. 
\end{acknowledgement}

\begin{suppinfo}
Extraction of polariton properties from the reflectance measurements, fit of the polariton dispersion, calibration of air gap in the experiment, estimation of polariton propagation lengths, extraction of exciton radiative lifetime, angle-resolved photoluminescence measurements, field distribution of the fundamental waveguide mode in the presence of SIL.

Supplementary video: evolution of angle-resolved reflectance maps with the change of the SIL--sample air gap. 
\end{suppinfo}

\bibliography{chemso}
\end{document}